# Cavity buildup dispersion spectroscopy

A. Cygan[1,*], A. J. Fleisher[2], R. Ciuryło[1], K. A. Gillis[2], J. T. Hodges[2], D. Lisak[1]

[1]Institute of Physics, Faculty of Physics, Astronomy and Informatics, Nicolaus Copernicus University in Toruń, Grudziadzka 5, 87-100 Torun, Poland

[2]National Institute of Standards and Technology, 100 Bureau Drive, Gaithersburg, Maryland 20899, USA

*agata@fizyka.umk.pl

January 23, 2020

*Measurements of ultrahigh-fidelity absorption spectra can help validate quantum theory, engineer ultracold chemistry, and remotely sense atmospheres[1-4]. Recent achievements in cavity-enhanced spectroscopy using either frequency-based dispersion[5] or time-based absorption[6] approaches have set new records for accuracy with uncertainties at the sub-per-mil level. However, laser scanning[5] or susceptibility to nonlinearities[6] limits their ultimate performance. Here we present cavity buildup dispersion spectroscopy (CBDS) in which the dispersive frequency shift of a cavity resonance is encoded in the cavity's transient response to a phase-locked non-resonant laser excitation. Beating between optical frequencies during buildup exactly localizes detuning from mode center, and thus enables single-shot dispersion measurements. CBDS yields an accuracy limited by the chosen frequency standard, a speed limited by the cavity round-trip time, and is currently 50 times less susceptible to detection nonlinearity compared to intensity-based methods. The universality of CBDS shows promise for improving fundamental research into a variety of light-matter interactions.*

Highly accurate models of light-matter interactions are important for fundamental studies of molecular hydrogen[1,7], tests of molecular structure calculations[8,9], modelling of planetary atmospheres[3,10], and the development of advanced spectroscopic databases[11]. The latter application is crucial for improvements in remote sensing and measurements of variations in greenhouse gas concentrations at 0.1% uncertainty levels required to better predict changes in Earth's climate[4]. To develop and test these models using first-principles approaches, accurate experimental techniques are required. Cavity mode-dispersion spectroscopy[12] (CMDS) is one such technique recently developed to meet these challenges[5]. While CMDS yields absorption spectra entirely in terms of measured optical frequency shifts[12], the shifts are not read out instantaneously. Consequently, the CMDS technique is susceptible to drifts. Although frequency-agile rapid scanning (FARS) spectroscopy[13] provides single-shot acquisition of local absorption limited only by the cavity response time, it is an intensity-based technique and inherently susceptible to nonlinearities in the detection system[6]. Therefore, each of these established techniques in ultrasensitive absorption spectroscopy has a critical weakness.

Here, we present cavity buildup dispersion spectroscopy (CBDS): an accurate, phase-sensitive measurement of cavity mode frequency that can be implemented on time scales that are substantially shorter than the cavity buildup time. In CBDS, a phase-locked laser beam is instantaneously injected into a high-finesse cavity followed by observation of the transient transmitted signal. The net response involves optical interference between the excitation and the transient cavity fields, with the latter field always occurring at the local cavity resonance frequency[14] and in opposition to the former field. Thus, measurement of the resulting heterodyne beat frequency provides the local cavity mode position. In practice, absorption-induced dispersion within the resonator leads to changes in the measured mode-by-mode beat frequencies, which yield dispersion spectra. We demonstrate that CBDS measurements can be made on timescales equal to or significantly less than the buildup duration and show the method to be relatively immune to nonlinear response in the detection system.

Recently reported techniques in condensed-phase sensing use micro-resonators with optical quality factors of $Q \leq 10^8$ to readout dispersive signals on nanosecond timescales[15,16]. As described by Yang et al.[17], those approaches clearly leverage heterodyne[18] and rapidly swept laser[19,20] cavity ring-down readout concepts developed decades ago to reduce technical noise and improve sensitivity. Collectively, the prior works in dispersive sensing utilize decaying signals which occur after the cavity driving field is effectively extinguished. Consequently, and unlike in the present work, dispersive micro-resonator sensors do not probe the temporal region associated with cavity buildup—a regime which includes two synchronous fields of interest here: a non-resonant laser-driven field with rapidly changing amplitude and the cavity's transient response to this field which always occurs at the cavity resonance frequency.

Uniquely, we demonstrate here dispersion spectroscopy performed in the transient buildup regime. The CBDS approach requires the sequential injection of discrete and arbitrarily detuned laser fields which are optically phase-coherent with the resonant cavity field[21]. Using high-$Q$ (~$10^{11}$), macroscopic-length (~1 m) resonators we demonstrate acquisition times as short as 3 µs (Fig. 1b)—far from the fundamental lower limit set by the cavity round-trip time and Nyquist-Shannon sampling criteria (e.g., $2t_{rt} = 4nL/c$ ~ 1 ns for an effective geometric length of 0.1 m)—and therefore achieve a new measurement paradigm without sacrificing the ability to study a wide range of dynamic phenomenon. In addition, we establish the immunity of CBDS to common nonlinearities and biases which may occur with conventional intensity-based cavity-enhanced spectroscopy methods[6,22]. We note that the interrogation of consecutive cavity modes via buildup signals to measure broadband phenomena such as molecular absorption spectra has not been previously considered.

The general concept and experimental realization of CBDS are illustrated in Fig. 1. The CBDS method utilizes a single-frequency light source that is phase-locked to the optical resonator. We use a double-polarization phase-locked laser scheme[5,13] where one linear polarization of laser light is phase-locked to a cavity mode while the orthogonal polarization (having a well-controlled, constant frequency detuning $\nu_{MW}$ from the locking point) is used for non-resonant excitation of the measured cavity mode (Fig. 1a). Additionally, the cavity length is stabilized to prevent thermal drift of the comb of modes over time scales >1 s. The buildup signal is initiated after rapidly switching on the frequency-detuned probe beam at the measurement mode, and the locking beam remains on during the entire cavity pumping process (but does not contribute to the CBDS signal thanks to a polarization-dependent optical filter and/or offset locking). Coherent averaging in the time-domain of repeated events is readily achieved because of the tight phase-locking scheme. The Fourier spectrum of the beating signal appearing in the transmitted light allows determination of the frequency detuning $\delta\nu_{meas}$ of the cavity resonance with respect to the probe beam frequency $\nu_P$. Finally, frequency agile rapid scanning for fast spectral acquisition is achieved by adjusting the detuning frequency $\nu_{MW}$ using a high-bandwidth (~20 GHz) electro-optic modulator[5,13].

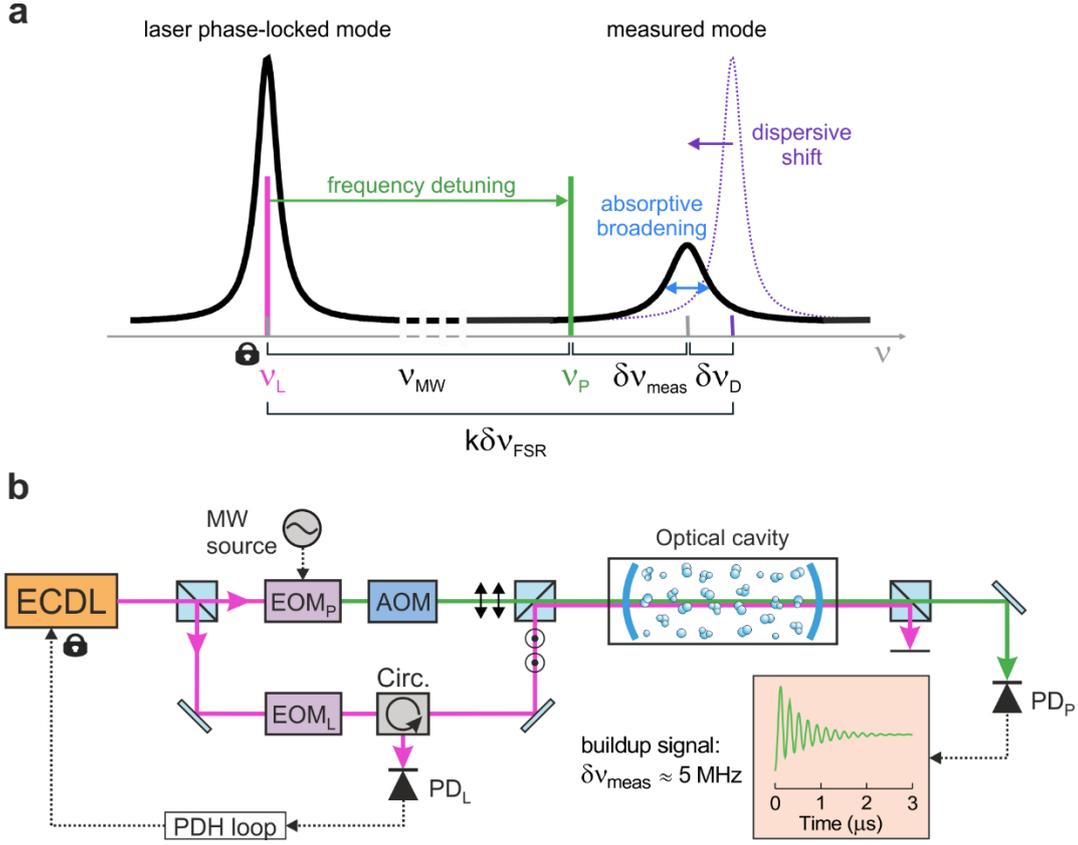

**Figure 1 Cavity mode localization and schematic of CBDS apparatus. (a)** The laser frequency $\nu_L$ is phase-locked to a $TEM_{00}$ cavity mode, which is a local resonance of the cavity transmission spectrum indicated by the thick black line. An orthogonally polarized beam with frequency $\nu_P$, is detuned from $\nu_L$ by frequency $\nu_{MW}$ and excites another $TEM_{00}$ mode shifted by dispersion. For non-resonant excitation, an oscillation on the transmitted buildup signal with a frequency $\delta\nu_{meas}$ corresponds to heterodyne beating between the non-resonant driving field and the resonant transient response of the cavity. For a given mode $k$ relative to the locking point and cavity free spectral range $\delta\nu_{FSR}$, the dispersive shift $\delta\nu_D$ of the cavity mode can be retrieved from $\delta\nu_{meas}$. **(b)** Schematic of the CBDS experiment. A broadband electro-optic modulator $EOM_P$, driven by the microwave source MW, rapidly detunes the ECDL beam with frequency $\nu_P$ from the locking point. An acousto-optic modulator AOM prevents cavity excitation by the carrier frequency. A photodiode $PD_P$ records the buildup signal. Here the experimental signal for $\delta\nu_{meas} = 5$ MHz and time interval 3 µs is shown. $EOM_L$, Circ. (circulator), $PD_L$: elements in the Pound-Drever-Hall phase-locking loop.

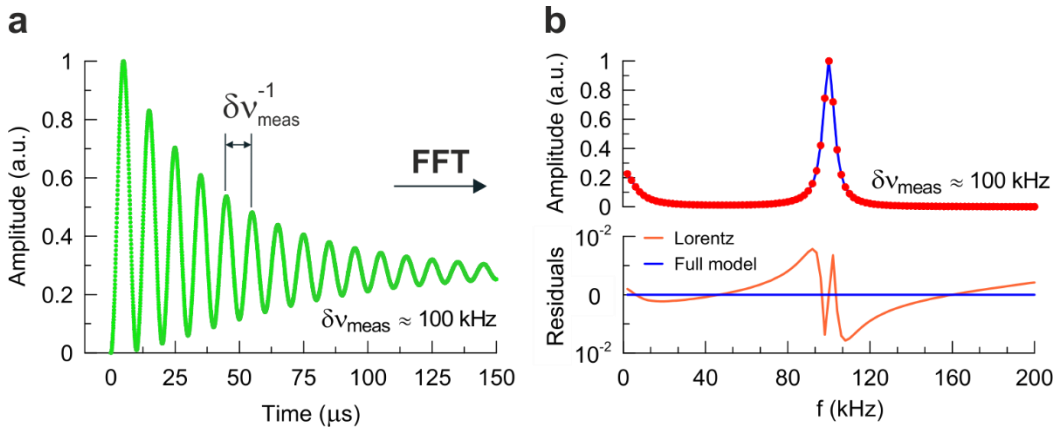

**Figure 2 Transient cavity response to single-mode, non-resonant excitation. (a)** Simulated buildup signal for $\delta\nu_{meas} = 100$ kHz and a driving field switch-on time of 50 ns. For clarity, the initial 150 µs of the signal is shown.

**(b)** The FFT spectrum of the buildup signal from the panel (a) and residuals from fits of Lorentzian (orange) and our asymmetric (blue) model.

In Fig. 2a a simulated buildup signal is shown. We developed a physically justifiable model for the transient cavity response to non-resonant single-mode excitation which is based on summation of delayed replicas of the driving field with a phase shift growing during each round-trip time[23]. Fourier transformation of our model reproduces the fast-Fourier transform (FFT) of the measured transmitted light intensity with $10^{-7}$ accuracy and works $10^5$ times better than a single Lorentzian function, which does not capture the predicted asymmetries (Fig. 2b). Even greater improvement is observed for retrieved values of $\delta\nu_{\text{meas}}$ leading to $10^{-9}$ accuracy. The relation $k\delta\nu_{\text{FSR}} - (\nu_{\text{MW}} + \delta\nu_{\text{meas}})$ yields the dispersive shift $\delta\nu_D$ of the cavity resonance relative to the locking point frequency $\nu_L$ (Fig. 1a). Here, $\delta\nu_{\text{FSR}}$ is the cavity free spectral range (FSR) corresponding to cavity conditions outside the molecular resonance with potential contribution from the broadband intracavity dispersion, and $k$ is the integer number of modes between the locking point and the measured cavity mode.

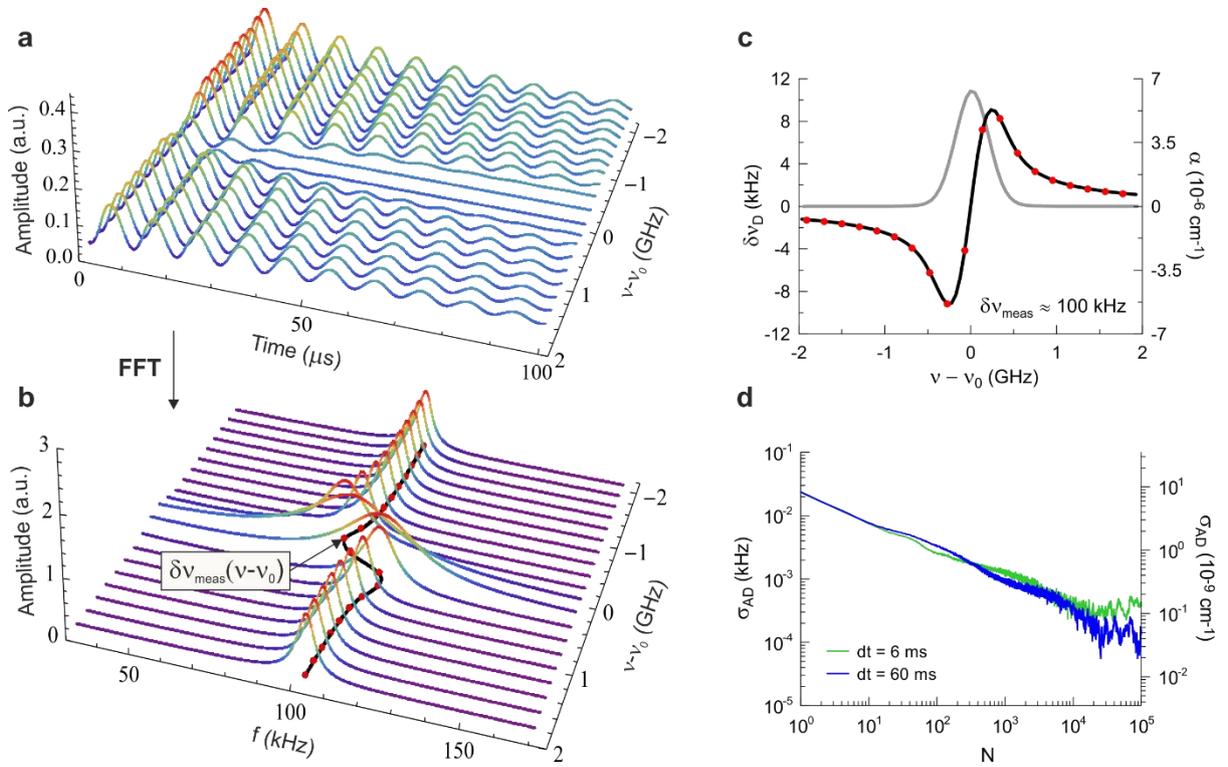

**Figure 3 Cavity buildup dispersion spectroscopy.** **(a)** Buildup signals recorded for cavity modes around the R23 CO molecular line centered at $\nu_0 = 192193.3341$ GHz and having an intensity of $8.056 \times 10^{-25}$ cm/molec. Each signal is an average over 150 scans recorded one by one and separated by 1.4 ms (0.2 ms for the buildup signal and 1.2 ms for the subsequent ring-down decay and change of microwave generator frequency $\nu_{\text{MW}}$). **(b)** The FFT spectrum of buildup signals from the panel (a). Absorptive broadening and dispersive shift of the FFT peaks (cavity modes) are visible. Fitting of the frequency-domain spectrum provides accurate positions $\delta\nu_{\text{meas}}(\nu - \nu_0)$ of cavity modes within the molecular line. **(c)** The dispersive spectrum $\delta\nu_D(\nu - \nu_0)$ reconstructed from the FFT peak frequencies shown in the panel (b). The absorptive spectrum (gray line) is calculated from the dispersive spectrum by the use of a complex-valued line shape model involving the real and imaginary components of the resonant susceptibility. **(d)** Allan deviations of $\delta\nu_{\text{meas}}$ versus the number of samples for two different time intervals between samples. The vertical axis is expressed in frequency shift and corresponding absorption units.

For each cavity resonance, a single buildup signal was recorded at a new value of $\nu_{\text{MW}}$ (Fig. 3a). Corresponding FFT spectra, presented in Fig. 3b, show absorptive and dispersive changes in the width and position of cavity modes within the frequency range of the measured molecular line. Measured dispersive shifts $\delta\nu_D(\nu - \nu_0)$, determined for each cavity mode, were used to reconstruct the purely frequency-based complex-valued line shape of a CO transition of central frequency $\nu_0$ (Fig. 3c). Allan deviation plots of $\delta\nu_{meas}$, Fig. 3d, demonstrate excellent stability of the frequency measurement in the CBDS experiment yielding an equivalent absorption coefficient detection limit less than $3 \times 10^{-11}$ cm$^{-1}$, corresponding to a detection limit of ~100 mHz. Moreover, consistent with the rapid phase-sensitive nature of the measurement, a short-term 20-Hz sensitivity to cavity resonance shifts was obtained in 400 µs.

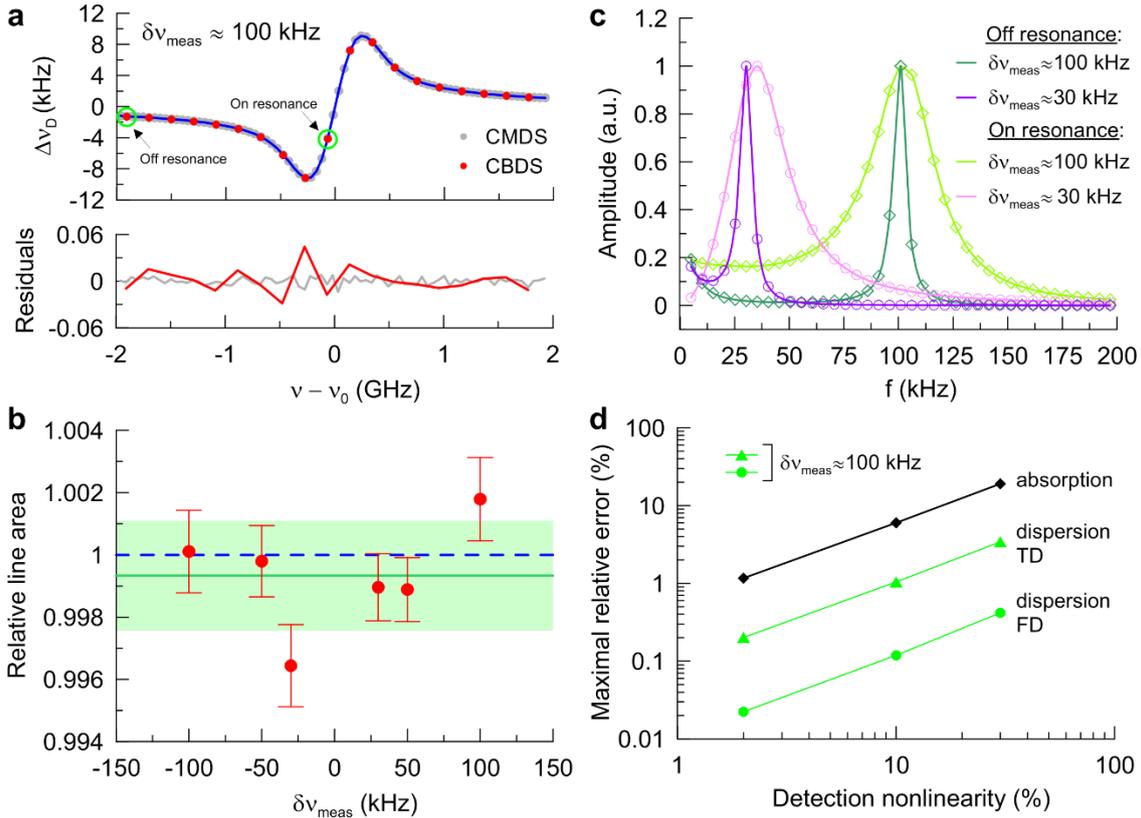

**Figure 4 Accuracy of CBDS spectra.** (**a**) Comparison of experimental CBDS (red points) and CMDS (gray points) dispersive spectra. Below are residuals from the fit with the Hartmann-Tran profile. (**b**) Line areas obtained from CBDS relative to CMDS as a function of detuning $\delta\nu_{meas}$. The green shaded region corresponds to ±1σ standard deviation of results. (**c**) The FFT spectra of buildup signals measured for $\delta\nu_{meas} \approx 30$ kHz and 100 kHz, for cavity modes located on and off the molecular resonance - green circles on the panel (a). Visible asymmetry of the FFT peaks increases with smaller detuning $\delta\nu_{meas}$. Our model for the Fourier spectrum reproduces well the experimental data enabling accurate determination of cavity mode position. (**d**) Maximum relative systematic errors of absorptive and dispersive spectra plotted as a function of quadratic detection nonlinearity.

In Figs. 4a-b we demonstrate excellent agreement between CO spectra and peak areas obtained from CMDS and CBDS experiments. We found that that systematic differences between line areas determined from CBDS and CMDS methods are only 0.07 % on average, with a standard deviation of 0.17 %. These observations indicate that the accuracy of these first CBDS measurements are already similar to the most accurate techniques currently available[5,6]. We emphasize that this level of agreement requires the proper frequency-domain modelling (Fig. 4c) of the transmission signals.

To quantify the effect of detection system nonlinearity on the accuracy of CBDS, we simulated buildup signals and assumed nonlinear quadratic or power-law deviations from linearity of the signal amplitude. Conventional cavity ring-down spectroscopy (CRDS) absorption spectra also were analyzed in the same fashion. Given the same degree of assumed nonlinear response, the maximum relative errors in CBDS analyzed in the time-domain (TD) and frequency-domain (FD) were found to be independent of detuning $\delta\nu_{\text{meas}}$ and 6 and 50 times smaller, respectively than those acquired using CRDS, see Fig. 4d. For the frequency-domain CBDS case, maximum errors were 0.02% for realistic non-linearities at the 2% level[6,22].

CBDS achieves high accuracy through precise measurement of the cavity resonance frequencies, and accurate modeling and fitting of the buildup signals in the frequency domain. The generality of our field-based method enables applications to dynamic cavity-enhanced sensing throughout the electromagnetic spectrum, making the method amenable to the analysis of intermode[24] and multiplexed[25] buildup signals with detuned local excitation fields. When dynamic events are not of interest, the tightly locked optical scheme allows for coherent time-domain averaging of CBDS signals, and therefore ultrahigh precision with minimal data storage and no dead time. We see the potential of CBDS for improving the accuracy of fundamental and atmospheric absorption spectroscopy studies as well as metrological applications which to date have depended exclusively on intensity-based experiments. The rapidity and sensitivity of CBDS should also render it useful in fast biological processes and single-particle spectroscopy. Moreover, CBDS methods can be readily extended to broadband spectroscopic techniques using an optical frequency comb, which will open new possibilities for high-accuracy measurements in this field[26].

The CBDS aligns well with general efforts to express physical quantities in terms of frequency[27]. Molecular spectra entirely measured in terms of cavity resonance frequencies can be easily referenced to the atomic frequency standard. CBDS will result in robust SI-referenced uncertainties and will greatly facilitate interlaboratory comparisons of data. In this context, we see clear applications of CBDS e.g. to Doppler thermometry[28] as well as to a new gas pressure standard currently being developed which is based on precisely measuring the dispersive shifts of optical cavity modes[29]. Also, recent nondestructive detection of Rydberg atoms based on cavity dispersion[30] indicates the potential of CBDS in terms of both speed and accuracy for determination of atomic population.

**Methods**

*Transient cavity response to single-mode, non-resonant excitation*

Consider a conventional, linear optical cavity formed by two mirrors having intensity reflectivity $R$ and separated by a distance $L$. The cavity is filled with an intracavity gas medium described by an absorption coefficient $\alpha$. We define an effective mirror reflectivity and round-trip time of the empty cell as $R_{\text{eff}} = Re^{-\alpha L}$ and $t_r = 2nL/c = 1/\delta\nu_{\text{FSR}}$, respectively, where $c$ is the speed of light in vacuum, $n$ is the refractive index of absorptive medium and $\delta\nu_{\text{FSR}}$ is the cavity free spectral range. Let us consider excitation of the cavity by light electric field $E_i(t) = \epsilon(1 - e^{-\Gamma_0 t})e^{i\omega_c t}$ characterized by an arbitrary angular frequency $\omega_c$ and amplitude $\epsilon$. Here, we assume a finite switch-on time $\tau_0 = \Gamma_0^{-1}$ of the electric field. The time response of the cavity $E_{\text{out}}(t)$ can be calculated at a given time by summing the contribution of finite number of $M$ passes in the cavity realized up to this moment[23]

$$E_{\text{out}}(t) = (1 - R)e^{-\alpha L/2} \sum_{m=0}^{M} R_{\text{eff}}^m E_i(t - mt_r). \qquad (1)$$

We assumed that $t = 0$ corresponds to the moment when the first transmitted field contribution leaves the cavity. Further expansion of Eq. (1) leads to the sum of two finite geometric series

$$E_{\text{out}}(t) = A(\alpha)e^{i\omega_c t}\left[\sum_{m=0}^{M}\left(R_{\text{eff}}e^{-i\delta\omega t_r}\right)^m - e^{-\Gamma_0 t}\sum_{m=0}^{M}\left(R_{\text{eff}}e^{\Gamma_0 t_r}e^{-i\delta\omega t_r}\right)^m\right], \quad (2)$$

where the factor

$$A(\alpha) = \epsilon(1-R)e^{-\alpha L/2} \quad (3)$$

describes modification of the electric field amplitude after the first pass through the cavity. In transition from Eq. (1) to Eq. (2) we replaced the expression $e^{-i\omega_c t_r}$ by $e^{-i\delta\omega t_r}$, since the electric field angular frequency can be rewritten as $\omega_c = 2\pi(N\delta\nu_{\text{FSR}} + \delta\nu)$, where $N$ is the cavity mode number and $\delta\omega = 2\pi\delta\nu$ is detuning of the light angular frequency from the cavity mode center. Small values of round-trip time $t_r$ allow us to replace the discrete time values $mt_r$ by the continuous quantity $t$ and consequently use integrals instead of sums in Eq. (2)

$$E_{\text{out}}(t) = A(\alpha)e^{i\omega_c t}t_r^{-1}\left[\int_0^t e^{(t_r^{-1}\ln(R_{\text{eff}})-i\delta\omega)t'}dt' - e^{-\Gamma_0 t}\int_0^t e^{(t_r^{-1}\ln(R_{\text{eff}})+\Gamma_0-i\delta\omega)t'}dt'\right]. \quad (4)$$

Here, for the convenience of calculations we expressed $R_{\text{eff}}^{t/t_r}$ as $\exp[t/t_r\ln(R_{\text{eff}})]$. Evaluation of Eq. (4) leads to the final complex-valued expression for the electric field leaving the cavity

$$E_{\text{out}}(t) = E_{\text{out}}^0(t)\left[e^{i\omega_c t} - D(t)e^{-\Gamma_q t}e^{i(\omega_c - \delta\omega)t}\right], \quad (5)$$

where

$$E_{\text{out}}^0(t) = B(1 - Ce^{-\Gamma_0 t}), \quad (6)$$

$$B = \frac{A(\alpha)}{t_r}\frac{1}{\Gamma_q + i\delta\omega}, \quad (7)$$

$$C = \frac{\Gamma_q + i\delta\omega}{\Gamma_q - \Gamma_0 + i\delta\omega}, \quad (8)$$

$$D(t) = \frac{1 - C}{1 - Ce^{-\Gamma_0 t}}, \quad (9)$$

and $\Gamma_q = -t_r^{-1}\ln(R_{\text{eff}})$ describes the width (HWHM) of $q$-th cavity mode. It can be easily shown that $(2\Gamma_q)^{-1}$ is the conventional intensity-based time constant of a light decay measured in cavity ring-down spectroscopy. The structure of Eq. (5) illustrates how the transient field tends to oppose the driving field and giving rise to interference between the two fields at frequencies $\omega_c$ and $\omega_q = \omega_c - \delta\omega$, respectively. Taking the real part of the field defined by Eq. (5) gives

$$\text{Re}\{E_{\text{out}}(t)\} = |E_{\text{out}}^0(t)|\cos(\omega_c t + \varphi_1) - |E_{\text{out}}^0(t)D(t)|e^{-\Gamma_q t}\cos[(\omega_c - \delta\omega)t + \varphi_2], \quad (10)$$

where

$$\varphi_1(t) = \text{arctg}\{\text{Im}[E_{\text{out}}^0(t)]/\text{Re}[E_{\text{out}}^0(t)]\}, \quad (11)$$

$$\varphi_2(t) = \text{arctg}\{\text{Im}[E_{\text{out}}^0(t)D(t)]/\text{Re}[E_{\text{out}}^0(t)D(t)]\}. \quad (12)$$

In order to compute the intensity, we square the real-valued field (Eq. 10) and average all sinuosoidal terms over optical cycles to account for the finite detector bandwidth. Ignoring the sum frequency term occurring at optical frequencies, this operation yields the intensity exiting the cavity as

$$I_{\text{out}}(t) = I_{\text{out}}^0(t)\{1 + |D(t)|^2 e^{-2\Gamma_q t} - 2|D(t)|e^{-\Gamma_q t}\cos[\delta\omega t - \varphi(t)]\}. \quad (13)$$

that fully describes the shape of the buildup signal measured in the CBDS method. Here, the amplitude of the transmitted signal is defined as $I_{\text{out}}^0(t) = |E_{\text{out}}^0(t)|^2/2$. The intensity function given by Eq. (13)

exponentially approaches a constant value for long times and exhibits damped oscillations at the beat angular frequency, $\delta\omega$ between that of the excitation field, $\omega_c$ and the cavity resonance frequency, $\omega_q$. Note the occurrence of two characteristic rates, one at $2\Gamma_q$ equal to the familiar ring-down intensity decay rate, and the other at half of this value, which corresponds to the characteristic decay rate of the field amplitude.

In general, the amplitude $I_{\text{out}}^0(t)$, the factor $|D(t)|$ and the phase $\varphi(t) = \varphi_2(t) - \varphi_1(t) = \text{arctg}\{\text{Im}[D(t)]/\text{Re}[D(t)]\}$ in Eq. (13) are time-dependent functions. However, for small values of $\tau_0$ all these functions can be treated as time independent. In practice, for $\tau_0 = 50$ ns the expression $e^{-\Gamma_0 t}$ is of the order of $10^{-9}$ for $t > 1\mu s$, so it can be neglected in $I_{\text{out}}^0(t)$ and $D(t)$ leading to time-independent functions $I_{\text{out}}^0$ and $D$. For the limiting case $\tau_0 \to 0$ ($\Gamma_0 \to \infty$), which corresponds to an immediate switch-on of the incident light electric field, $I_{\text{out}}^0(t) \to |B|^2$, $D(t) \to 1$ and $\varphi(t) \to 0$ (because $C \to 0$).

*Spectrum model of the buildup signal*

For practical reasons mentioned above let us consider the buildup shape function $\Phi(t)$ approximating the ratio $I_{\text{out}}/I_{\text{out}}^0$ from eq. (13) in the form

$$\Phi(t) = 1 + D^2 \exp(-2\Gamma_q t) - 2D \exp(-\Gamma_q t) \cos(\delta\omega t - \varphi), \tag{14}$$

where $I_{\text{out}}^0$, $D$ and $\varphi$ are real, time-independent parameters corresponding to original time-dependent quantities $I_{\text{out}}^0(t)$, $|D(t)|$, $\varphi(t)$, respectively.

The Fourier transform $\mathcal{F}(\omega) = \int_0^\infty \Phi(t) \exp(i\omega t)\, dt$ leads to the formula

$$\mathcal{F}(\omega) = \frac{D^2}{2\Gamma_q - i\omega} - D\left[\frac{e^{-i\varphi}}{\Gamma_q - i(\omega + \delta\omega)} + \frac{e^{i\varphi}}{\Gamma_q - i(\omega - \delta\omega)}\right], \tag{15}$$

in which we have omitted the Dirac delta function term. The Fourier transform spectrum $I_{\text{out}}(\omega) = \mathcal{F}(\omega) \cdot \mathcal{F}^*(\omega)$ of the buildup signal is given as

$$I_{\text{out}}(\omega) = \frac{D^2}{(2\Gamma_q)^2 + \omega^2} + D^2\left[\frac{1}{\Gamma_q^2 + (\omega - \delta\omega)^2} + \frac{1}{\Gamma_q^2 + (\omega + \delta\omega)^2}\right] +$$
$$2D^2 \frac{(\Gamma_q^2 + \omega^2 - \delta\omega^2)\cos(2\varphi) + (2\Gamma_q \delta\omega)\sin(2\varphi)}{(\Gamma_q^2 + \omega^2 - \delta\omega^2)^2 + (2\Gamma_q \delta\omega)^2} -$$
$$2D^3 \left\{ \frac{[2\Gamma_q^2 + \omega(\omega + \delta\omega)]\cos(\varphi) + [2\Gamma_q(\omega + \delta\omega) - \Gamma_q \omega]\sin(\varphi)}{[2\Gamma_q^2 + \omega(\omega + \delta\omega)]^2 + [2\Gamma_q(\omega + \delta\omega) - \Gamma_q \omega]^2} + \right.$$
$$\left. \frac{[2\Gamma_q^2 + \omega(\omega - \delta\omega)]\cos(\varphi) - [2\Gamma_q(\omega - \delta\omega) - \Gamma_q \omega]\sin(\varphi)}{[2\Gamma_q^2 + \omega(\omega - \delta\omega)]^2 + [2\Gamma_q(\omega - \delta\omega) - \Gamma_q \omega]^2} \right\}. \tag{16}$$

This function, apart from Lorentzian components, also contains symmetric, dispersive terms. We checked numerically that $I_{\text{out}}(\omega)$ given by Eq. (16), scaled by an amplitude parameter, reproduces the fast Fourier transform (FFT) spectrum of the buildup signal simulated from Eq. (13) with $10^{-7}$ accuracy. Here, the limiting factor is the accuracy of the FFT spectrum calculation caused by the finite sampling density of the buildup signal. We also found that adding a constant background parameter to Eq. (16) further improves the agreement between our model and FFT spectrum by more than a factor $10^3$.

*Effect of detection system nonlinearity on measurement accuracy*

To quantify how system nonlinearity affects the accuracy of measured spectra, we performed simulations of normalized buildup signals, $T(t)/T_{max}$, with their amplitudes multiplied by the function $y_1(t) = 1 - a[T(t)/T_{max}]$, where $a$ is a constant factor which scales the degree of nonlinearity and $T_{max}$ is the maximum amplitude of $T(t)$ for the whole spectrum (Fig. I). We also did analogous calculations to evaluate the sensitivity of conventional cavity ring-down spectroscopy (CRDS) absorption spectra to detector nonlinearity. We found a 0.1% - 4% systematic bias in the *y* axis of CRDS spectra and a 1% - 20% maximum systematic error for $2\% < a < 30\%$ (Fig. Ib). Here, detector nonlinearity of 2% corresponds to a realistic case[6,22]. For CBDS spectra retrieved from frequency-domain (FD) analyses of buildup signals, the maximum systematic error is up to 50 times smaller than that predicted for CRDS spectra and is independent of detuning $\delta\nu_{meas}$. The systematic bias of the FD CBDS spectra averages close to zero within the entire spectrum (Fig. Ia). We estimate sub-per-mil accuracy in the FD CBDS even when the detector nonlinearity is as high as 8%. Moreover, we noticed that the choice of fitting the buildup signals in the time or frequency domain has a large impact on the sensitivity to detector nonlinearity. Dispersive spectra obtained through time-domain (TD) analyses of buildup signals have a nonlinear susceptibility intermediate between the CRDS and the FD CBDS cases, i.e. 6 times lower by comparison to the CRDS absorption case. In the case of the TD CBDS spectrum, we also calculate a systematic bias of the *y* axis ranging from 0.01% - 0.2 % and we observed a slight asymmetry of TD CBDS spectra (Fig. Ia). As a further exploration of the effect of non-linearity, we assumed its power law response model has the form $y_2(t) = [T(t)/T_{max}]^a$, which in conventional CRD spectroscopy would lead to measured decay rates biased by the constant fractional amount, $a$. Notably, this type of nonlinearity would not be evidenced in the fit residuals of individual decay signals because the decay signals remain exponential in form. As can be seen in Fig. Ib, this type of nonlinearity augments the bias for both the CRDS and TD CBDS spectra, by comparison to $y_1(t)$, but yields nearly identical results for the FD CBDS case.

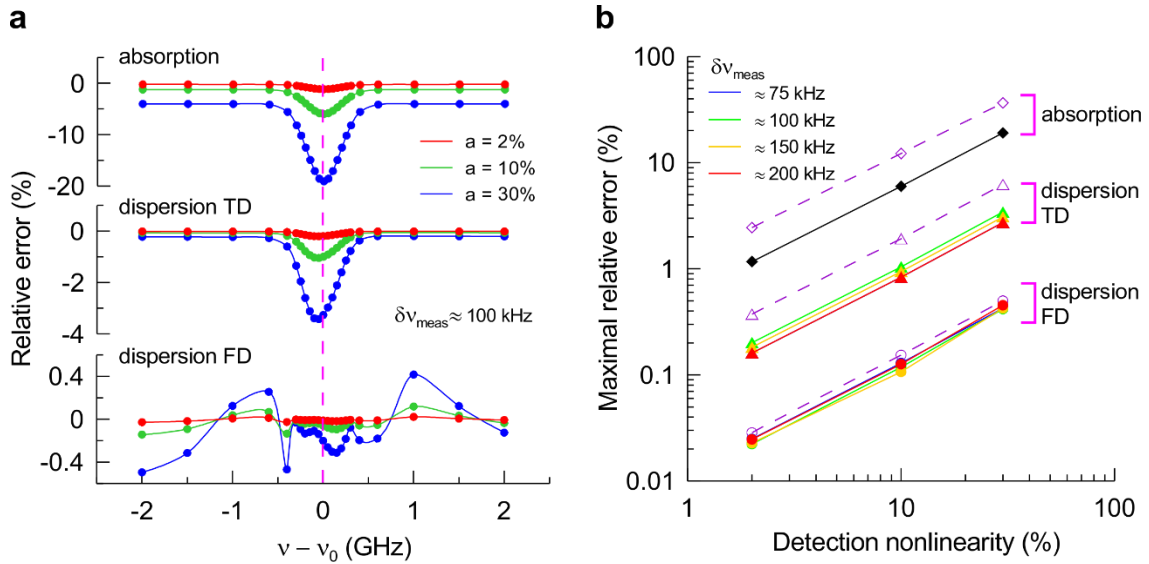

**Figure I Influence of detection system nonlinearity on the spectrum accuracy.** (a) Relative systematic differences between absorptive/dispersive spectra obtained from ring-down/buildup signals simulated with and without nonlinearity of the amplitude for corresponding nonlinear factors *a*: 2%, 10% and 30%. The buildup signals were analyzed in the time and frequency (as a power spectrum) domains leading to spectra denoted as "TD" and "FD", respectively. For dispersion, we chose a detuning $\delta\nu_{\text{meas}} \approx 100$ kHz, corresponding to $\approx 30$ cavity mode widths (HWHM). We modeled nonlinear distortion of the spectrum by multiplying the normalized time response of the cavity $T(t)/T_{\text{max}}$ (buildup and ring-down signals) by $y_1(t) = 1 - a[T(t)/T_{\text{max}}]$, where $T_{\text{max}}$ is the maximum amplitude of $T(t)$. (**b**) Solid lines - maximum relative systematic errors of absorptive and dispersive spectra from the panel (a) versus detection nonlinearity $y_1$ for various detunings $\delta\nu_{\text{meas}}$. Dashed purple lines - similar results when using the power-law model of detector nonlinearity given by $y_2(t) = [T(t)/T_{\text{max}}]^a$ and detuning $\delta\nu_{\text{meas}} \approx 100$ kHz.

## Acknowledgements

The research is part of the program of the National Laboratory FAMO in Toruń, Poland. The research was supported by the National Science Center, Poland project Nos. 2015/18/E/ST2/00585 and 2015/17/B/ST2/02115. AJF, JTH and KAG were funded by NIST.

## Author contributions

AC and JTH independently conceived the idea of cavity buildup dispersion spectroscopy. AC and DL designed the concept of the research, built the experimental setup and wrote software for experiment and data analysis used in the manuscript. AC and DL at NCU performed all measurements, data analysis, simulations and figures presented in the manuscript. Concurrent experiments performed at NIST by AJF also revealed the CBDS method. All authors contributed to developing the theory of cavity buildup spectroscopy. AC wrote the original draft of the manuscript, and all authors contributed to the final manuscript. DL at NCU and JTH at NIST coordinated the project.

## Competing interests

JTH, KAG and AJF are named inventors on NIST provisional U.S. patent application 62/951,200 "Laser apparatus and methods for performing transient intramode heterodyne cavity buildup spectroscopy", which covers in part a laser apparatus and methods for performing transient heterodyne spectroscopy in a resonator. The remaining authors declare no competing interests.